# Anthropogenic Hybridization at Sea: three evolutionary questions relevant to invasive species management

**Frédérique Viard[1], Cynthia Riginos[2] and Nicolas Bierne[3]**
*1 Laboratory AD2M, Station Biologique, Sorbonne Université, CNRS, Roscoff, France*
*2 School of Biological Sciences, The University of Queensland, St Lucia QLD, Australia*
*3 ISEM, Univ Montpellier, CNRS, EPHE, IRD, Montpellier, France.*

## Summary

Species introductions promote secondary contacts between taxa with long histories of allopatric divergence. Anthropogenic contact zones thus offer valuable contrasts to speciation studies in natural systems where past spatial isolations may have been brief or intermittent. Investigations of anthropogenic hybridization are rare for marine animals, which have high fecundity and high dispersal ability, characteristics that contrast to most terrestrial animals. Genomic studies indicate that gene flow can still occur after millions of years of divergence, as illustrated by invasive mussels and tunicates. In this context, we highlight three issues: 1) the effects of high propagule pressure and demographic asymmetries on introgression directionality, 2) the role of hybridization in preventing introduced species spread, and 3) the importance of postzygotic barriers in maintaining reproductive isolation. Anthropogenic contact zones offer evolutionary biologists unprecedented large scale hybridization experiments. In addition to breaking the highly effective reproductive isolating barrier of spatial segregation, they allow researchers to explore unusual demographic contexts with strong asymmetries. The outcomes are diverse from introgression swamping to strong barriers to gene flow, and lead to local containment or widespread invasion. These outcomes should not be neglected in management policies of marine invasive species.

# Introduction

Human-mediated translocations of species across oceans have been ever increasing especially since the onset of the twentieth century, due to growing international trade [1]. These biological introductions have complex consequences on the distribution of biological diversity. Introductions of non-indigenous species (NIS) inherently disrupt natural biogeographic barriers [2]. Thus, secondary contacts between previously allopatric species are important outcomes of biological introductions. Because species boundaries are often semipermeable, human-mediated translocation of species creates opportunities for hybridization and introgression [3] which, together with habitat disturbances, contribute to "anthropogenic hybridization" (for a review and references, see [4, 5]).

Such human-mediated secondary contacts create unique opportunities to examine the processes enhancing or preventing hybridization and introgression in detail and in real-time [6]. They offer replicated situations for investigating the strength of reproductive isolation barriers that accumulated during divergence in allopatry. These insights are of particular relevance for considering strongly differentiated yet incompletely isolated species that do not have the opportunity to meet at natural hybrid zones, such as species living in separate oceans or continents. For instance, Pacific-Atlantic congeners are historically separated by 3+ million years (MY) divergence in allopatry [7], yet many species native to the Pacific have been introduced to the North East Atlantic where they encounter their congeners. This history is typified by the Pacific sea squirt *Ciona robusta* (see Box 1). When these anthropogenic contact zones involve taxa with long histories of divergence in allopatry, they offer valuable contrasts to speciation studies in natural systems where past spatial isolations may have been brief and intermittent. These long allopatric species likely fall at the extreme of the speciation continuum and exhibit high levels of reproductive isolation. Following the rationale proposed by Hewitt [8] (i.e., hybrid zones as natural laboratories for evolutionary studies), anthropogenic hybridizations between such species thus provide *in situ* laboratories to examine the fate of co-occurring divergent genomes. These are also excellent opportunities to examine genomic changes i) in new environments for the introduced species, and ii) in novel genetic backgrounds for both the native and introduced genes (Box 2). Additionally, some recent secondary contact situations can illuminate our understanding of repeated periods of inter-taxa gene flow as a mixture between past and contemporary introgression events, both leaving co-existing footprints at the genome level ([9] and Box 1).

The outcomes of hybridization following human-mediated secondary contacts are diverse from extensive asymmetric introgression (introgression swamping), genome-wide admixture, to semi-permeable barriers to gene flow involving strong coupling between isolating loci [10]. The specific outcomes depend on the strength of concomitant endogenous and exogenous reproductive barriers [10]. In this context, secondary contacts can have important consequences for the fate of NIS. They may favor the sustainable establishment of NIS (for instance through adaptive introgression of "ready-to-use" alleles). They can sometimes lead to the emergence of novel species (i.e. homoploid hybrid speciation) or well identified hybrid genetic clusters, as reported particularly for terrestrial plant species [4, 11]. Human-mediated secondary contacts and subsequent hybridization thus have important implications for conserving local biota [12, 13].

Hybridization has attracted much attention in terrestrial NIS studies, especially in plants ([14] and references herein), but is rarely documented in marine ecosystems. Marine NIS are, however, numerous in coastal marine systems: for instance, about 1400 NIS are reported in European Seas [15]. Admixture and hybridization involving marine NIS are also likely to be common [16, 17], particularly for broadcast spawners ([18] and references herein). Intra-specific admixture, defined as the mixing between historically isolated but non-reproductively isolated lineages, has been addressed elsewhere ([16] and references therein]. Yet, only a few hybridizing species with strong reproductive isolating mechanisms have been examined in marine systems and this will be the focus of this paper. Two of the best-studied empirical cases of secondary contacts are presented in Boxes 1 and 2, highlighting the sea squirts *C. intestinalis* (native to the North Atlantic) and *C. robusta* (native to the North Pacific) and similarly the blue mussels *Mytilus trossulus* (native to the North Pacific) and *M. galloprovincialis* (native to the Mediterranean and East Atlantic).

Demography and migration are pivotal factors influencing the evolution of hybrid zones, and the natural history attributes that characterize marine introduced animals will likely create distinctive demographic and dispersal syndromes. First, many marine NIS have very high reproductive rates, releasing hundreds to thousands of eggs or larvae during a reproductive season [19], or even during a single reproductive event (e.g., the invasive gastropod *Crepidula fornicata* [20]). Second, most marine organisms have a highly dispersive stage allowing rapid spread and potentially extensive gene flow. These inherent characteristics of marine NIS play out against human modified conditions where long distance and local transport associated with shipping and aquaculture can serve as ongoing introduction vectors and pathways. Similarly, restricted openings of ports and harbours might retain many larvae creating local high abundance of species inhabiting these artificial habitats [21]. Together these biological and anthropogenic factors can yield very high propagule pressure, defined as the number of individuals multiplied by the number of introduction events, which is evidenced by genetic diversity that is often similar or even higher in NIS than that in native range populations [17]. High densities of NIS and natives living sympatrically in novel environments also may dampen prezygotic isolation, as habitat preferences in natural environments would normally prevent interspecific gametes from direct interactions.

These properties, hardly ever encountered in terrestrial animals, can influence the outcome of human-mediated secondary contacts. Although anthropogenic marine hybridizations are starting to receive attention, emerging themes across studied taxa highlight dynamics that contrast to terrestrial systems. We outline three pressing questions regarding how human-mediated secondary contacts affect the trajectory of marine invasive species spread. We then highlight their resultant implications for NIS management and marine conservation of coastal areas.

**Could propagule pressure counter-balance demographic assymetry?**

Propagule pressure had been shown to be a good proxy for invasive success [22] and can be very high for marine NIS [17]. However, even if numerous propagules are introduced, census population sizes of NIS are often minute relative to that of native congeners in the receiving community. This difference in population size is particularly important at the initial stage of introduction, leading to a demographic imbalance between non-native and native species and consequently affecting introgression when interbreeding occurs. If the NIS establishes itself and grows in abundance (or propagule pressure is sufficiently high), hybridization dynamics can subsequently reverse.

There are three reasons why introgression should mainly occur from the native into the NIS genome, especially during two introduction stages: i) the initial phase when the NIS is still in low abundance, and ii) the propagating phase before the NIS-native hybrid zone halts at an equilibrium point. First, a barrier to gene flow depends on population densities and should proceed from the dense into the sparse population [23]. Second, later along the invasion trajectory, introgression should travel from the established receding taxon into the propagating one, with hybrids being dominated by backcrosses toward the non-native parental genome in the invasion wavefront ([24] and references herein). This prediction is corroborated in mussels by the slight but detectable introgression of native *M. trossulus* alleles into the invading *M. galloprovincialis* genome (Box 2). Third, given that the native genome is already adapted to the local environment whereas the NIS genome is encountering a new environment, the NIS could use "ready-to-use" adaptive alleles from the native species (i.e. adaptive introgression).

However, a hybrid zone can subsequently stabilize at a dispersal or an environmental boundary. This trapping of hybrid zones is theoretically expected when the barrier is maintained by intrinsic and/or extrinsic selection against hybrids ([25] and references therein). Once the hybrid zone is trapped, hybridization will enter a new phase and the direction of introgression may then reverse. By this time, the invasive population is likely to have become larger and denser; movement will be interrupted and introgression should proceed from the fittest (NIS) into the less fit (inbred native) taxon. If some portions of the native genome contain large numbers of deleterious mutations, these portions should be replaced by the NIS counterpart [26] (in other words, genetic rescue proceeds [27]).

Genetic pollution is often feared in invasion biology [28]. However, if there is a semi-permeable barrier between the native species and the NIS, hybrid zone theory predicts that deleterious alleles should be filtered out during the introgression process. Therefore, genetic pollution is a worry predominantly in a situation when the two interacting taxa are not isolated by strong reproductive isolation. For loosely isolated taxa, propagule pressure can induce a sufficient migration load (introgression of deleterious or locally maladapted alleles) into the native population, analogous to gene flow from hatchery reared to wild stocks, such as recorded for salmon [29]. In addition, when reproductive isolation is weak, the NIS genome hardly resists asymmetric introgression along the invasion front [24]. Indeed, genetic swamping (i.e., the admixture of the two genomes) is of concern in this situation. This is, for instance, possibly what is happening the Indo-Pacific sergeant major damselfish (*Abudefduf vaigiensis*) and its endemic congener, *A. abdominalis* in Hawaii [30]. Genetic swamping might also be the case between the two lineages of the green crab *Carcinus maenas* introduced in North East America, although their place along the population-species continuum is unclear [31]. For well-isolated species, the genetic effect of propagule pressure should mainly affect NIS populations. High propagule pressure would likely drive the genetic composition of the introduced population(s) back to the composition found in the native range, providing the spread of the invasion front is not too rapid and quickly halted by a dispersal boundary. This could be the case in ports of the North East Atlantic colonized by admixed "dock mussels" (hybrids between the native blue mussel *Mytilus edulis* and the introduced Mediterranean mussel *Mytilus galloprovincialis*; Box 2), which so far, have remained confined into ports [32] despite a likely continual influx of new propagules from shipping traffic.

**Genetic Allee effect, our ally against NIS expansion?**
The Allee effect, where population growth is slow or negative when population density is low (*demographic Allele effect*), is a cornerstone concept in invasion biology [33]. Therefore, successful introductions are expected to rely on high propagule pressure to push the population density of the founding population above the threshold for positive population growth ([22] and references herein). An Allee effect could also slow, or even halt, range expansion at the leading front when migration is low [34]. A reduced growth rate at the leading front also allows for maintaining high genetic diversity due to migration from the core population (pushed wave behavior, [35]). In an inspiring model, Mesgaran et al. [36] showed that hybridization allows NIS to escape the Allee effect, when mate-limited, by relying on the population dynamics of the native species (hybridization rescue). However, the demographic advantage of hybridizing with congeners in an already established population is expected to work only when reproductive isolation is weak.

The evolution of intrinsic barriers with bi-stable dynamics enhances reproductive isolation, which in turn creates a *genetic Allee effect* [37]. When at low density, NIS will fail to find conspecific partners and will produce unfit hybrids when mating with natives, such that the fitness of NIS is low for genetic reasons rather than for demographic reasons. As with the demographic Allee effect, a genetic Allee effect requires a sufficiently high propagule pressure in order to initiate a spreading wave (i.e., pushed wave dynamics of bi-stable variants, [38]). In [36], taxa can be considered as partially reproductively isolated when the compatibility parameter, $\beta$ is below 1, then hybridization opposes colonization rather than facilitating it. Following initial successful introduction, for example in a place with few individuals of the native species, the subsequent spatial spread can easily and durably be halted by the first dispersal barrier or density trough encountered [39].

To date, there is no unified theory combining both demographic and genetic Allee effects to predict invasion success and spread. However, the available theory suggests that reproductively semi-isolated species should be efficiently confined when the hybrid zone, formed between the NIS and the native species stops moving and halts at the first dispersal barrier. We posit that this could be the case for dock mussels in ports [32] (Box 2), for the present contact between *M. galloprovincialis* and *M. trossulus* in California [40] (Box 2), and for the cyclic ephemeral breakthroughs of *Ciona robusta* within *C. intestinalis* populations in marinas of the English Channel [41, 42] (Box 1). A hybrid zone identified between two lineages of the marine snail *Stramonita haemastoma* in eastern Spain could also be explained by the trapping of an invading hybrid zone at the first dispersal barrier it encountered [43]. Often the status of non-native taxa is defined by geographical isolation rather than specific evidence of reproductive isolation. The observation of fast introgression and genetic swamping (as in the aforementioned Hawaiian sergeant major damselfishes [30] and green crab ecotypes [44]) would indicate that there is minimal reproductive isolation and thus admixture proceeds largely unimpeded for these species.

**What is the relative importance of pre- vs. postzygotic barriers?**

A noteworthy aspect of marine introductions is that they frequently occur in artificial habitats, such as marinas, ports, and aquaculture facilities. The proliferation of built structures and artificial hard substrates (better known as "coastal hardening" or "ocean sprawl") substantially contributes to the establishment of marine sessile NIS [21]. Besides being points-of-entry for NIS, these artificial habitats are not surrogates of natural rocky reefs: they are particular habitats hosting numerous NIS alongside native species. The consequences of these novel niches and species interactions deserve further scrutiny especially from an evolutionary perspective.

Because of their biotic (e.g., specific assemblages) and abiotic (e.g., substrates, pollutants) specificities, these artificial habitats will likely select for species pre-adapted to these particular human-made habitats [45], in contrast to natural habitats where native species have evolved in a specific ecological and environmental context. They are also likely to constitute more homogeneous environments at a global scale. Similarly, to terrestrial plants [14], we can thus hypothesize that extrinsic prezygotic barriers will play a less important role than postzygotic barriers in preventing hybridization between species in these human-made artificial habitats. In addition, the observation of introgression between broadcast spawners with very large molecular divergence (at least as evidenced by mtDNA *COI*) is consistent with few or weak intrinsic prezygotic barriers in externally fertilizing marine invertebrates [18], although this hypothesis has not yet been explicitly tested. Investigations of harbor populations of *Ciona* spp., for example, show that the native and non-native species live in closed syntopy, overlap in spawning and recruitment time [41], and F1's can be artificially created [46] suggesting minimal prezygotic isolation (Box 1).

Importantly, even if postzygotic barriers are particularly effective following secondary-contacts, this does not preclude extrinsic prezygotic barriers evolving such as via reinforcement following secondary contacts [47]. Ports and marinas are indeed heterogeneous habitats at small scales. For instance, floating pontoons differ from fixed substrate pillars or seawalls in their species assemblages and NIS contribution [48]. Subtle ecological niche differentiation might thus contribute to reinforce the isolation, via habitat preference, between native and non-native species. An important caveat to the hypothesis of reduced prezygotic isolation is that experimental crosses to gauge prezygotic isolation are rarely undertaken because they are difficult to execute. Even more infrequently is postzygotic isolation estimated from experimental crosses, given the challenges in rearing planktonic larvae to settlement and for more than one generation. More commonly, reproductive isolation is inferred from observed genotypic frequencies of field populations, where high frequency of F1 hybrids but low frequency of backcrosses may indicate Dobzhansky-Muller interactions expressed in the second generation of hybridization (typifying mussel hybrid zones: see [49] and Box 2).

In general, postzygotic barriers are expected to scale with species divergence [50], as illustrated by *Tigriopus* copepods [51] or *Strongylocentrotus* sea urchins [52]. In line with this expectation, reproductive isolation is substantial between *C. robusta* and *C. intestinalis* (Box 1) and similarly high between *M. galloprovincilis* and *M. trossulus* (divergence time of 3.5 MY, Box 2), whereas reproductive isolation is negligible for taxa with recent divergence times (e.g., *M. galloprovincialis* X *M. planulatus* [53]; damselfish in Hawaii [30]; green crabs [44]), and intermediate between *M. galloprovincialis* and *M. edulis* with a mixture of heterosis and hybrid breakdown due to multigenic interactions [9] (divergence time of ~2.5 MY [38]). In the context of marine anthropogenic hybridization, the observations made in *Ciona* spp. suggest that prezygotic isolation may be less important in shaping evolutionary and ecological outcomes than postzygotic isolation.

**Marine invasive species policies will benefit from speciation studies**

Speciation is a gradual process during which barriers to gene exchanges are accumulating. However, as pointed out in the Introductory paper of this issue [54], speciation does not necessarily follow a linear progression. Sudden change may arise. One of the most effective barriers to genetic exchanges, namely spatial barriers, are disappearing instantaneously with human-mediated secondary contacts. It is thus not surprising that "*genetic pollution*", "*hybrid swarm*", "*extinction by hybridization*" are phrases commonly encountered in the biological invasion literature. Marine anthropogenic hybridizations are no exception. And yet, so far, NIS have largely been overlooked in conservation planning for Marine Protected Areas [55].

Evolutionary studies can provide valuable insights regarding the future outcomes of anthropogenic hybridization. They have for instance documented that anthropogenic hybridizations influence the fate of both native and non-native species. NIS can threaten native species when hybridization occurs between two congeners, such as shown in the native grey-ducks in New-Zealand following the introduction of the mallard duck (e.g., [56]). Anthropogenic hybridizations can lead to the emergence of novel invasive species, as illustrated by the notorious allopolyploid cordgrass *Spartina anglica* [57]. Between compatible taxa, hybridization can also facilitate invasion by reducing mate limitation and overcoming demographic Allee effects, as suggested in *Cakile* plant invaders [36]. However, focusing here on marine systems, we also have shown that hybridization may dampen NIS spread, through the establishment of hybrid zones and genetic Allee effects. This is particularly true for species that have developed strong, yet incomplete, reproductive isolation mechanisms. Determining the speciation stage and strength of the isolating barriers between NIS and native species may indicate whether hybridization can oppose invasion, and thus contribute to estimating invasion risks.

Controlling propagule pressure might also be an effective strategy for limiting the spread of NIS genotypes and alleles, even if introductions have already occurred. Reducing NIS propagule pressure, through effective control of introduction pathways and vectors including ballast water or aquaculture trade, will increase the asymmetry in hybridization success, so that if "genetic pollution" should occur, it would be mostly directed towards the invader. Reducing propagule pressure might be particularly important when the native congeners are endangered or rare, hence when the demographic imbalance between native and non-native species is low. Many marine introductions are reported in anthropogenic habitats, a situation best explained by these structures being hubs of connectivity enduring high propagule pressure along with some kind of pre-adaptation to similar habitats globally distributed across oceans. In this context, NIS and natives might have non overlapping ecological niches in their natural (native) home ranges, but this could be altered completely in artificial habitats. Using ecological niche modelling based on observation in natural habitats in the native range may thus be misleading for inferring future distributions of NIS and associated prezygotic isolation in no analogue microenvironments. It is however noteworthy that for species at late speciation stage, reinforcement processes through prezygotic mechanisms could potentially evolve in the introduced range. Altogether, any mechanisms aimed at reducing NIS density, containing NIS in anthropogenic habitats, and minimizing their escape, should be particularly beneficial for controlling their expansion.

Mirroring issues related to protection of endangered species [58], considering hybridization in the context of biological invasions may also spark difficult debates, such as how to categorize anthropogenic vs. natural hybridization. This latter debate is particularly acute in a context of on-going climate change, another category of human-driven range expansion [59], which may lead to extinction though hybridization (although not yet documented [60]). The population-species continuum, a cornerstone of speciation theories, and an outcome of recent genomic studies [50], also creates conundrums for conservation objectives [61]: should we focus on protecting genes or phenotypes or fuzzily defined species? Similar questions are raised with invasive species management, as illustrated by recent studies of the invasive green crab *Carcinus maenas*. Two lineages, geographically separated in the native range and both introduced in North East America, were shown to be genome-wide divergent and to hybridize [44]. Introgression occurred in a few locations, suggesting some reproductive isolation mechanisms at play between so-called "ecotypes" [44], which may differ by their ecology (e.g., winter sea water temperature preferences), therefore questioning that they will spread more or less [44, 62]. An opposite case is illustrated by the cupped oysters *Crassostrea gigas* and *C. angulata,* both introduced in Europe, which showed little genome-wide divergence [63]. These case studies, and those cited in the preceding sections, showcases how pairs of native vs. non-native marine species are spread along the population-species continuum. Because the outcomes of these secondary contacts depend on the speciation stage, evolutionary studies investigating reproductive isolation mechanisms can provide valuable information regarding the future outcomes of anthropogenic hybridization involving marine NIS.

Natural hybridization is not well understood and anthropogenic hybridization is a new area of research. It offers replications of recent contacts that may provide fruitful information on the hybridization process. To date we can foresee hybridization as probably being as much a problem as a solution to invasion [13]. It is legitimate to fear the negative effects of "genetic pollution", but this concern should be balanced by the positive effects of "genetic rescue" [27]. In a rapidly changing world, targeting evolutionary processes including speciation in conservation planning seems a worthwhile approach [61, 64]. Hybridization is an evolutionary process with important consequences for genetic and phenotypic diversity. Anthropogenic hybridization does not necessarily cause negative outcomes and may be important in some situations for halting invasive species spread. We should neither favor nor fear it.

## Acknowledgments

The authors are thankful to their present and former PhD students especially Sarah Bouchemousse, Christelle Fraisse, Marine Malfant, Iva Popovic, Alexis Simon for their hard work on *Mytilus* and *Ciona* species that largely contributed to the opinions presented in this paper. These ideas also arose from the ANR HySea project (No. ANR-12-BSV7-0011).

**BOX 1. The sea squirts *Ciona* spp.: a case study of secondary contact with cyclic and ephemeral breakthrough of the introduced species**

Ascidians are a major component of NIS, particularly in artificial habitats such as harbors, docks and piers, where many of them appear as cosmopolitan invaders. This is notably the case of *Ciona robusta* (formerly known as *C. intestinalis* type A, before an in-depth taxonomic revision [65]). Presumably native to the North West Pacific, *C. robusta* has been introduced to the North East and South Pacific, and to the North and South Atlantic (see map in [66]). The North East Atlantic (especially the English Channel and the Bay of Biscay) is the only area where *C. robusta* had been described living in sympatry with its native congener *C. intestinalis* (formerly known as *C. intestinalis* type B), from which it has probably diverged >3 MY ago [67]. The introduced species display variations in abundance over seasons and years, with episodic breakthrough [18, 41].

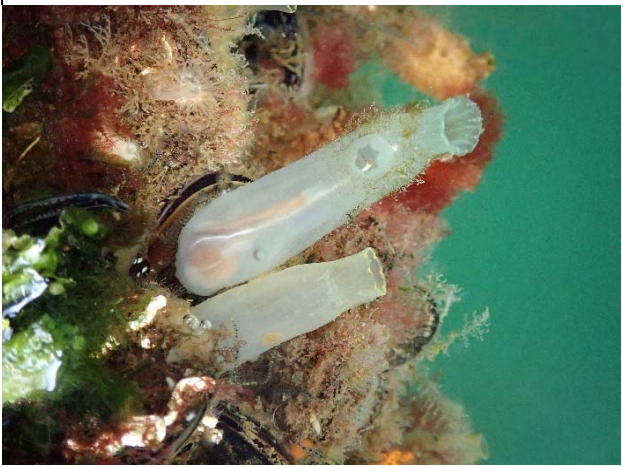

**Figure. The sea squirts *Ciona robusta* (top specimen) and *C. intestinalis* (bottom specimen) are living in closed syntopy in harbours and marinas of the English Channel. They are distinguishable based on subtle morphological criteria. One of them is the red pigmentation of the terminal papillae of the vas deferens in *C. robusta* (as visible in this picture).**
**Photo credit: Laurent Lévêque**

Interestingly, based on transcriptome sequences, an approximate Bayesian computation framework shows that the two species have historically hybridized (although when and in which geographic location is unknown) such that there is a distinct footprint of past introgression in both species [67]. Such multiple contacts may characterize other marine NIS and also highlight the necessity for detailed historical analyses before presuming that all shared polymorphisms arise solely from contemporary hybridization [42].

Comprehensive field and experimental studies show that the two species produced gametes synchronously with juveniles recruited at the same time (twice a year for both species), and are easily crossed in the lab with F1 hybrids showing no signs of outbreeding depression [41, 46, 68]. Although reproductive isolating mechanisms are expected to scale with divergence times, hybridization has thus been documented at a late stage of speciation following human-mediated species translocation. However, the use of an ancestry-informative SNPs panel show that introgression is negligible and only few hybrids had been detected in localities where the two species live in syntopy [42]. Additional experiments documented that F1 hybrids produce less sperm, and F2 backcross hybrids display a reduced survival as compared to parental species (M. Malfant & F. Viard, unpublished data). Altogether, the empirical patterns observed, notably the crossing experiments, suggest postzygotic selection in agreement with the prediction of a multigenic determinism based on Fisher's geometric model [49].

**BOX 2. The blue mussel *Mytilus galloprovincialis*: replicated introductions could support comparative studies**

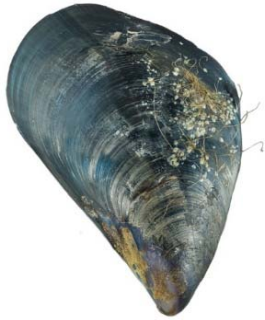

The Mediterranean blue mussel, *Mytilus galloprovincialis* has been introduced to many different regions in the world with the greatest evidence for substantive ecological impacts in South Africa where there is no native congener. First observed in western South Africa in 1979, *M. galloprovincialis* now encompasses over 2000 km of coastline, where it greatly restructured the intertidal community (e.g., [69]), with at least one documented extinction of a local bivalve [70]. Aside from South Africa, all other receiving communities for *M. galloprovincialis* host native congeners, without remarkable ecological influence.

Photo credit: Jessa Thurman

Where there is a native congener, hybridization has been reported between *M. galloprovincialis* and the native species. The hybrid zone between *M. galloprovincilis* and the native *Mytilus trossulus* on the East Pacific coast of North America shows a broad scale cline from north (*M. trossulus* genotypes) to south (*M. galloprovincialis* genotypes) [40]. Mussels collected prior to 1900 from the Los Angeles region and further north have exclusively *M. trossulus* mitotypes, indicating that *M. galloprovincilis* displaced *M. trossulus* from its southern range [71]. The sequential spread from the likely point of introduction, possibly pausing at dispersal barriers encountered, could be due to the introduced species being adapted to warmer temperatures [72], or even being intrinsically fitter than the native species whatever the environment. *M. galloprovincialis* also hybridizes with *M. trossulus* in the West Pacific where connectivity and adaptation to temperature might also play a role: in Hokkaido Island, Japan the distribution of parental types from 2004-2006 surveys aligned well with major currents and their associated temperature regimes, where *M. trossulus* were most common in the cooler waters of the east [73].

Endogenous reproductive isolation mechanisms seem to limit introgression between *M. galloprovincialis* and *M. trossulus*. In the East Pacific, hybrids are present at low frequencies but primarily consist of F1's and first generation backcrosses [40]. Also, consistent with the concept that numerical dominance of the native species and spatial propagation of the invasion front into the native range could cause introgression of native alleles into the introduced species' background, a recent genomic study found a small number of *M. galloprovincialis* backcross individuals and no *M. trossulus* backcross individuals along the East Pacific coast of North America [40]. Many hybrids also have anomalous mitochondrial genome compositions by sex with an overall excess of female mussels [74]. Likewise, in the Hokkaido Island hybrid zone, very high proportions (>50%) of hybrid mussels failed to produce mature eggs or sperm [73].

*M. galloprovincialis* has also been introduced to Chile [75], Kerguelen Islands [76], and Australia and New Zealand [53, 77, 78]. Intriguingly, a new study finds very recent (<50 years) introduction of Mediterreanean type *M. galloprovincialis* into several Atlantic harbours accompanied by strikingly parallel clinal hybrid zones where hybrids with a majority Mediterreanean type *M. galloprovincialis* background predominate in the harbors [32]. These "dock mussels" thus seem confined to harbors conversely to native genotypes (*M. edulis* and Atlantic *M. galloprovincialis*) that are found outside the harbors [32]. These various contact zones deserve further investigation, as examining replicated situations can foster our understanding of rules and identify common isolating mechanisms.